# IEEE Copyright notice:

# Critical Clearing Time Sensitivity for Inequality Constrained Systems

Chetan Mishra, *Member, IEEE*, Anamitra Pal, *Member, IEEE*, and Virgilio A. Centeno, *Senior Member, IEEE*

*Abstract*—From a stability perspective, a renewable generation (RG)-rich power system is a *constrained* system. As the quasi-stability boundary of a constrained system is structurally very different from that of an unconstrained system, finding the sensitivity of critical clearing time (CCT) to change in system parameters is very beneficial for a constrained power system, especially for planning/revising constraints arising from system protection settings. In this paper, we derive the first order sensitivity of a constrained power system using trajectory sensitivities of fault-on and post-fault trajectories. The results for the test system demonstrate the dependence between ability to meet angle and frequency constraints, and change in power system parameters such as operating conditions and inertia.

## I. Introduction

As opposed to the *traditional* approach of tripping renewable generation (RG) offline during disturbances seen at the point of common coupling (PCC), the *modern* approach is to make them "ride through" such conditions. This has become necessary because systems with significant RG penetration could collapse due to loss of equilibrium [1], if a large quantity of such generation was lost at the time of need. That being said, RGs cannot be made to ride through *every* possible scenario. Therefore, ride through curves were devised in the form of time dependent voltage and frequency limits at the PCC of these generators, violation of which resulted in their tripping. Thus, a power system with high penetration of such generators could be seen as a dynamical system constrained to satisfy the ride through constraints [2]. The loss of stability phenomenon in constrained power systems is not only limited to loss of synchronism/voltage collapse but also involves the trajectory violating certain constraints which further results in undesirable structural changes in the system.

Critical clearing time (CCT) refers to the maximum time that can be taken to clear the fault and still retain stability. Usually, there is a monotonic relationship between fault clearing time and chances of instability and therefore it is used as a metric for computing transient stability margin. It is also desirable to understand the impact of system parameters on transient stability. For example, sensitivity of CCT to Q injection for a fault resulting in in tripping of large amounts of RGs due to low voltage ride through violations could help identify effective locations for new dynamic VAR resources to minimize such occurrences. Other parameters, such as system inertia (whose reduction is a growing concern with displacement of conventional generators by inverter based RG), also have significant impacts on CCT [3].

In the past, brute force approaches that relied on numerical integration were proposed for CCT sensitivity computation. Ayasun [4] reduced the multi-machine system to a single machine infinite bus system to evaluate the sensitivities; however, such an approach did not capture important phenomena of multi-machine systems. Chiodo and Lauria [5] used linear regression to understand the mapping between logarithm of CCT and loading. Nguyen [6] and Laufenberg [7] computed sensitivity of angle and speed trajectory in the post-fault phase w.r.t. fault clearing time since stable and unstable trajectories have significantly different $\omega$ limit sets. The most recent relevant work in this area is by Dobson [8] where the sensitivity of stable manifold of controlling unstable equilibrium point (CUEP) is used in conjunction with fault-on trajectory sensitivity to estimate the sensitivity of CCT to parameter changes. His derivation is for unconstrained ordinary differential equation (ODE) systems and an extension was proposed for differential algebraic equation (DAE) systems under the assumption that semi-singular surface is not contained inside the stability region (SR). In this paper, building upon Dobson's work, we derive CCT sensitivities for inequality constrained dynamical systems.

## II. Stability of Constrained Systems

The system being considered in this work is defined by the following state equation,

$$\dot{x}_{n\times 1} = f(x,p)_{n\times 1} \quad (1)$$
$$h(x,p)_{n_h\times 1} > 0$$

The first vector equation defines the evolution of states $x$ and the second one defines a feasibility region with boundary given by $\{x | (\prod_k h_k(x,p)) = H(x,p) = 0\}$. The system being studied is a function of parameter $p$ but the stability properties will be discussed with value of $p$ fixed. It is also clear from the above equation that the constraints do not have any impact on the system dynamics. A stable trajectory for such a system is defined as one that converges to a desired stable equilibrium point (SEP), $x_s$, and does not intersect the infeasible region $\{x | h_k(x,p) \leq 0 \ \exists k \in [1, n_h]\}$. Venkatsubramaniam [9] characterized the stability boundary for DAE systems with equality constraints. The instability phenomenon for such systems comprised of singularities and loss of synchronism, and involves hitting a bifurcation of the algebraic system, usually arising from incomplete modeling [10]. Loparo [11] extended this work to characterize the stability boundary of DAE systems with inequality constraints. In this section, we will extract it for systems with no algebraic constraints.

## A. Transformed Unconstrained System and Pseudo EPs

The traditional notion of an unstable trajectory is one that grows significantly apart from its stable counterparts due to the stability boundary having critical points. The feasibility boundary plays a defining role in the stability boundary structure of constrained systems since crossing it is also treated as an instability. However, it is usually devoid of points impacting the system dynamics making an unstable trajectory difficult to fit the above notion. As such, the constrained system given by (1) can be transformed to an equivalent unconstrained system [11] as shown in (2). It is important to mention here that this equivalent system will be used in this paper for understanding and distinguishing between the important points on the stability boundary of constrained systems, and not for deriving the CCT sensitivities.

$$\dot{x}_{n\times 1} = H(x,p)_{1\times 1} \times f(x,p)_{n\times 1} \quad (2)$$

In (2), all the inequality constraints are multiplied together making the first term on the RHS, $H(x,p)$, a scalar function. Normally, multiplying $H(x,p)$ with the vector field, $f(x,p)$, will only change the length and not the direction of the $f(x,p)$ vector. However, when any individual feasibility constraint is violated, $H(x,p)$ becomes negative, thus reversing $f(x,p)$. A closer look at the new vector field shows that the points on the feasibility boundary now also serve as EPs of this system, which we will refer to as *pseudo EPs*, denoted by $x_H$, to distinguish them from the original system's EPs, $x_e$. We now linearize the equation to understand the nature of pseudo EPs.

$$\Delta\dot{x} = \begin{bmatrix} \frac{\partial H}{\partial x_1}f_1(x,p) & \cdots & \frac{\partial H}{\partial x_n}f_1(x,p) \\ \vdots & \ddots & \vdots \\ \frac{\partial H}{\partial x_1}f_n(x,p) & \cdots & \frac{\partial H}{\partial x_n}f_n(x,p) \end{bmatrix} \times \Delta x + H(x,p)\times\frac{\partial f}{\partial x}\times \Delta x \quad (3)$$

The second term on the right of the above equation becomes 0 since $H(x,p) = 0$ for pseudo EPs. The connected components of $x_H$ represented by $\aleph_{x_H}$ is a $(n-1)$ dimensional manifold and thus, the state matrix in the first term on the RHS has $(n-1)$ eigenvalues equal to 0. Thus, the only non-zero eigenvalue equals the trace of this matrix given by $\sum_{i=0}^{n}\frac{\partial H}{\partial x_i}f_i(x,p) = \dot{H}(x,p)$. Therefore, a pseudo EP is stable (called $x_H^s$), if $f(x,p)$ points towards the feasibility boundary, and unstable (called $x_H^u$), if it points away, where the feasibility boundary serves as local center stable and center unstable manifold, respectively. Other important groups of point(s) which lie on the separating boundary between these two types of pseudo EPs are semi-saddle points (called $x_H^0$), in which case $f(x,p)$ is tangential as shown in Figure 1.

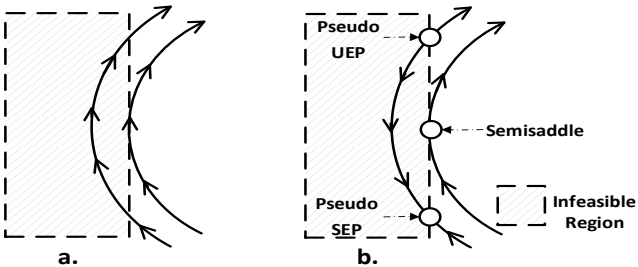

**Figure 1 Pseudo EPs a. Original System Dynamics (Eq. (1)); b. Transformed System Dynamics (Eq. (2))**

The connected components of semi-saddle points $\aleph_{x_H^0}$ has a maximal dimension of $(n-2)$ since it is defined by two equality constraints $\{x|H(x,p)_{1\times 1}=0, \dot{H}(x,p)_{1\times 1}=0\}$. $\aleph_{x_H}$ serves as the local tangent hyperplane to the stable manifold $W^s(\aleph_{x_H^0})$ which thus has the maximal dimension of $(n-1)$ with $\aleph_{x_H^0}$ serving as its boundary.

## B. Characterization of Quasi-Stability Boundary

The assumptions to be satisfied for the stability results are:

(A1) All original system EPs and periodic orbits on the stability boundary must be hyperbolic.

(A2) $W^s(x_e)$ and $W^s(x_H)$ must intersect transversally with $W^u(x_e)$ and $W^u(x_H)$. However, $W^s(\aleph_{x_H^0})$ and $W^u(\aleph_{x_H^0})$ must not be transversal for the same $\aleph_{x_H^0}$.

(A3) Any trajectory on the stability boundary must converge to one of the EPs or periodic orbits on the boundary.

The stability boundary of a generic nonlinear system can be very complex and can include truncated fractal structures. Therefore, from a practical perspective, the quasi-stability region [12] is a practical SR as its boundary is actually the boundary of closure of the stability region, $A(x_s)$, which can be written as $\partial\bar{A}(x_s)$. It has been shown that for constrained systems, this comprises of stable manifolds of type 1 original system UEPs denoted by $x_e^u$, type 2 periodic orbits, and $(n-2)$ dimensional $\aleph_{x_H^0}$ whose unstable manifolds intersect the SR. It may also include unstable portions of the feasibility boundary, $\aleph_{x_H^u}$. The results section shows the distinct characteristics of SRs of constrained systems.

## III. SENSITIVITY DERIVATION

### A. Overview

In this section, we analyze the impacts of small variations in $p$ on CCT. A critical fault-on trajectory for a given fault is one that intersects the stability boundary of the constrained post-fault system, $\partial A(x_s^{post})$. Without any loss of generality, the parametric quasi-stability boundary has maximal dimension $n$ and is of the form $S(x,p) = 0$. The intersection of parametric fault trajectory and stability boundary exists under parameter changes if they intersect transversally [13]. For this to happen, the tangent space of the parametric fault trajectory at the intersection point (which is also the vector field) should not be spanned by the tangent bundle of the parametric stability boundary. Thus, for the same change in $p$, CCT would be changed such that the new state at the fault clearing time lies on the new stability boundary.

In order to achieve this, we need to derive the sensitivity of the state value at the time of fault clearing, denoted by $x_{cl}$, for the fault-on trajectory, $\varphi^{fault}(x_0, t, p)$, and the associated relevant portion of the quasi-stability boundary it intersects. Now, there are three structurally distinct portions of the quasi-stability boundary viz. feasibility boundary itself, stable manifold of semi-saddle pseudo EPs, and stable manifolds of type 1 UEPs of the unconstrained system. Depending on the mode of loss of stability of a given critical fault trajectory, the appropriate sensitivity must be calculated. Here, we will present the derivations for the first two types since a formulation for the third type was already done in [6] and [8]. In the following sections, $p$ is assumed to be scalar. The sensitivity is computed for a given critical trajectory having CCT of $t_{cr}$, state vector value of $x_{cr}$, and $p = p_0$.

### B. Sensitivity of the State Value at Fault Clearing ($x_{cl}$)

Let us assume that the starting point of the fault-on trajectory $x_0^{fault}$ lies on a single dimensional manifold $l(x_0^{fault}, p) = 0$. Now, clearing time $t_{cl}$ is another parameter and therefore $x_{cl}$ lies on a manifold of maximal dimension two. Calculating sensitivity of $x_{cl}$, we get,

$$\left.\frac{\Delta x_{cl}(x_0, t_{cl}, p)}{\Delta p}\right|_{x_0^{fault}(p_0), t_{cr}, p_0} = M_1 \times \frac{\Delta x_0^{fault}}{\Delta p} + M_2 \times \frac{\Delta t_{cl}}{\Delta p} + M_3 \quad (4)$$

Where,

$M_{1(n \times n)} = \left.\frac{\partial \varphi^{fault}(x_0, t, p)}{\partial x_0}\right|_{x_s^{pre}(p_0), t_{cr}, p_0}$, $M_{2(n \times 1)} = \left.\frac{\partial \varphi^{fault}(x_0, t, p)}{\partial t}\right|_{x_s^{pre}(p_0), t_{cr}, p_0} = \left.f^{fault}(x, p)\right|_{x_{cr}, p_0}$, $M_{3(n \times 1)} = \left.\frac{\partial \varphi^{fault}(x_0, t, p)}{\partial p}\right|_{x_s^{pre}(p_0), t_{cr}, p_0}$

The sensitivity of $x_0^{fault}$ is given by,

$$\frac{\Delta x_0^{fault}}{\Delta p} = M_{4(n \times 1)} = -\left[\left.\frac{\partial l(x, p)}{\partial x}\right|_{x_0^{fault}(p_0), p_0}\right]^- \times \left.\frac{\partial l(x, p)}{\partial p}\right|_{x_0^{fault}(p_0), p_0} \quad (5)$$

Substituting (5) in (4), we get,

$$\left.\frac{\Delta x_{cl}(x_0, t_{cl}, p)}{\Delta p}\right|_{x_0^{fault}(p_0), t_{cr}, p_0} = M_1 \times M_4 + M_2 \times \frac{\Delta t_{cl}}{\Delta p} + M_3 \quad (6)$$

Usually in stability studies it is assumed that a fault trajectory starts from the parametric pre-fault system's SEP, i.e. $x_0^{fault} = x_s^{pre}(p)$ and $l(x, p) = f^{pre}(x, p) = 0$ for hyperbolic EPs. There are two ways how this could happen viz. it intersects $W^s(x_e^{post}$ or $\aleph_{x_H^0}^{post})$, or it intersects the feasibility boundary.

### C. Sensitivity of Combined Feasibility Boundary of Fault and Post-Fault System

Two ways in which the sustained fault trajectory can directly lose stability is by intersecting the (i) feasibility boundary of the fault-on system which is stable w.r.t. $f^{fault}(x, p)$ denoted by $\aleph_{x_H^S}^{fault}$, or (ii) feasibility boundary of the post-fault system which is unstable w.r.t. $f^{post}(x, p)$ denoted by $\aleph_{x_H^u}^{post}$. These two components can be combined together by multiplication, i.e. $\{x | H^{fault}(x) \times H^{post}(x) = 0\}$, to get the combined boundary. It should be kept in mind that there should not be constraint functions present in both $H^{fault}(x)$ and $H^{post}(x)$ as it may make $H^{comb}(x)$ positive definite. Now, the sensitivity of any point on the feasibility boundary is given by,

$$M_5 \times \frac{\Delta x_{cl}}{\Delta p} = M_6 \quad (7)$$

Where, $M_{5(1 \times n)} = \left.\frac{\partial H^{comb}(x, p)}{\partial x}\right|_{\varphi^{fault}(x_s^{pre}(p_0), t_{cr}, p_0), p_0}$, $M_{6(1 \times 1)} = -\left.\frac{\partial H^{comb}(x, p)}{\partial p}\right|_{\varphi^{fault}(x_s^{pre}(p_0), t_{cr}, p_0), p_0}$

Substituting (7) in (6) gives the sensitivity of CCT if the mode of loss of stability is direct intersection with the feasibility boundary. This is shown in the equation below.

$$\frac{\Delta t_{cl}}{\Delta p} = [M_5 \times M_2]^- \times (M_6 - M_5 \times (M_1 \times M_4 + M_3)) \quad (8)$$

### D. Sensitivity of Post-Fault Trajectory's End Point

The stability boundary could also contain $W^s$ of some EPs or pseudo EPs. That is, the stability boundary could contain a surface of adjacent trajectories all having the same $\omega$ limit set (type - 1 UEP or connected component of semi-saddle pseudo EPs) [14]. Since a critical fault trajectory is a one-dimensional manifold, it would be intersecting the stability boundary at a single point. Therefore, we will only focus on the emerging post-fault trajectory from that point, which we refer to as the *critical post-fault trajectory*. For a given critical post-fault trajectory to retain its criticality on variation of $x_{cl}$ due to change in $p$, the emerging post-fault trajectory must still lie on the new stability boundary. This can be guaranteed if its end point has the same $\omega$ limit set as the one whose $W^s$ forms the new stability boundary portion. Thus, the first step is to compute the sensitivity of the post-fault trajectory's end point $x_{end}^{post}$ which can be done in a manner similar to the fault-on system. If the given critical fault trajectory on clearing the fault takes time $T$ to reach the associated limit set with the value of state vector being given by $x_T = \left.x_{end}^{post}(x_{cl}, t, p)\right|_{x_{cr}, T, p_0}$, then,

$$\left.\frac{\Delta x_{end}^{post}(x_{cl}, t, p)}{\Delta p}\right|_{x_{cr}, T, p_0} = O_1 \times \frac{\Delta x_{cl}}{\Delta p} + O_2 \times \frac{\Delta t}{\Delta p} + O_3 \quad (9)$$

Where,

$O_{1(n \times n)} = \left.\frac{\partial \varphi^{post}(x_0, t, p)}{\partial x_0}\right|_{x_{cr}, T, p_0}$, $O_{2(n \times 1)} = \left.\frac{\partial \varphi^{post}(x_0, t, p)}{\partial t}\right|_{x_{cr}, T, p_0} = \left.f^{post}(x, p)\right|_{x_T, p_0}$, $O_{3(n \times 1)} = \left.\frac{\partial \varphi^{post}(x_0, t, p)}{\partial p}\right|_{x_{cr}, T, p_0}$

Substituting (6) in (9), we get,

$$\left.\frac{\Delta x_{end}^{post}(x_{cl}, t, p)}{\Delta p}\right|_{x_{cr}, T, p_0} = [O_1 \times M_2 \quad O_2] \times \begin{bmatrix} \frac{\Delta t_{cl}}{\Delta p} \\ \frac{\Delta t}{\Delta p} \end{bmatrix} + O_3 + O_1 \times (M_1 \times M_4 + M_3) \quad (10)$$

We will now derive the sensitivity of the different types of manifolds on which $\Delta x_{end}^{post}$ must lie to remain critical.

### E. Sensitivity of Stable Manifold of Semi-Saddle Pseudo EP

The connected component of semi-saddle pseudo EPs of the parametric post-fault system belongs to the set: $\{x | H^{post}(x, p) = 0, \frac{\partial H^{post}(x, p)}{\partial x} \times f^{post}(x, p) = 0\}$. If the mode of instability is observed as the post-fault trajectory eventually intersecting the feasibility boundary, $x_{end}^{post}$ should lie on this component. Calculating the sensitivity, we get,

$$O_4 \times \left.\frac{\Delta x_{end}^{post}(x_{cl}, t, p)}{\Delta p}\right|_{x_{cr}, T, p_0} = O_5 \quad (11)$$

Where, $O_{4(2 \times n)} = \begin{bmatrix} \left.\frac{\partial H^{post}(x, p)}{\partial x}\right|_{x_T, p_0} \\ \left.\frac{\partial \left[\frac{\partial H^{post}(x, p)}{\partial x} \times f^{post}(x, p)\right]}{\partial x}\right|_{x_T, p_0} \end{bmatrix}$ and

$$O_{5(2 \times 1)} = -\begin{bmatrix} \left.\frac{\partial H^{post}(x, p)}{\partial p}\right|_{x_T, p_0} \\ \left.\frac{\partial \left[\frac{\partial H^{post}(x, p)}{\partial x} \times f^{post}(x, p)\right]}{\partial p}\right|_{x_T, p_0} \end{bmatrix}$$

Combining (11) and (10), we get CCT sensitivity as,

$$\begin{bmatrix} \frac{\Delta t_{cl}}{\Delta p} \\ \frac{\Delta t}{\Delta p} \end{bmatrix} = [O_4 \times [O_1 \times M_2 \quad O_2]]^{-1} \times (O_5 - O_4 \times (O_3 + O_1 \times (M_1 \times M_4 + M_3)) \quad (12)$$

## IV. Overall Computation and Applications To Large Scale Systems

This section discusses the various computations involved in finding the sensitivity of CCT of a given fault to various parameter changes. Computationally tractable direct method for computing CCT for constrained systems is still a challenge because of the changes in nature of the stability boundary [2]. Therefore, for the constrained system under study, CCT as well as critical fault-on and post-fault trajectories are found using time domain simulation (TDS) for $p = p_0$ using Algorithm 1.

---

**Algorithm 1 – CCT and Critical Trajectory Computation using TDS for Constrained Systems**

i. **INITIALIZE** stable clearing time $t_{stable}$ and unstable clearing time $t_{unstable}$. $t_{unstable}$ is set to the time at which the sustained fault trajectory intersects the feasibility boundary.

ii. **SET** $t_{cl} = \frac{t_{stable} + t_{unstable}}{2}$. $\varphi^{fault}(x_s^{pre}(p_0), t_{cl}, p_0)$ is denoted by $x_{cl}$.

iii. **INITIALIZE** $t_1 = t_2 = \infty$. $x_T, T = null$

iv. Integrate the post-fault trajectory for a long enough time $T_{max}$.

v. **UPDATE** $t_1$ equal to time at which $H^{post}(x)$ crosses 0 or $H^{post}(x) \leq 1e - 5$.

vi. **IF** $t_1 \leq T_{max}$, $t_{unstable} = t_{cl}$.

vii. **IF** $\varphi^{post}(x_{cl}, T_{max}, p_0) \neq x_s^{post}(p_0)$, $t_{unstable} = t_{cl}$ and update $t_2$ to time where $\|f^{post}(x, p_0)\| \leq 1e - 3$ and acquires a local minimum value along post-fault trajectory.

viii. **IF** $\min(t_1, t_2) < \infty$, $T = \min(t_1, t_2)$, $x_T = \varphi^{post}(x_{cl}, T, p_0)$.

ix. **IF** $|t_{stable} - t_{unstable}| \geq 0.01$ **OR** $T = null$, **GOTO** ii.

x. **STOP**

---

The following items must be noted:

a. The transformed unconstrained system given in (2) for the post fault system can also be used for TDS. However, an adaptive step size is required for simulation since the time scale drastically varies with the value of $H(x)$ along a trajectory requiring a stiff system solver, which increases the TDS computation.

b. It is very difficult to precisely find the exact time at which a fault trajectory intersects the $W^s(CUEP)$. Therefore, we use the approach used in [15] for finding CUEP for gradient systems.

Besides the required Jacobian computations, the following trajectory sensitivities are also to be computed [16]:

i. Integrating the fault-on trajectory till $t_{cr}$ to compute $\frac{\partial \varphi^{fault}(x_0, t, p)}{\partial x_0}\bigg|_{x_s^{pre}(p_0), t_{cr}, p_0}, \frac{\partial \varphi^{fault}(x_0, t, p)}{\partial p}\bigg|_{x_s^{pre}(p_0), t_{cr}, p_0}$

ii. If the loss of instability is not direct intersection of the fault-on trajectory with the feasibility boundary, compute, $\frac{\partial \varphi^{post}(x_0, t, p)}{\partial x_0}\bigg|_{x_{cr}, T, p_0}, \frac{\partial \varphi^{post}(x_0, t, p)}{\partial p}\bigg|_{x_{cr}, T, p_0}$

When using the proposed approach on large scale systems, the main bottleneck is the computation of trajectory sensitivities. This can be overcome by using parallel programming and sparsity techniques as proposed in [17].

## V. RESULTS

In this section, we will use the following notations to denote the instability phenomenon: **1**: fault trajectory directly intersects the feasibility boundary, **2**: post-fault trajectory intersects the feasibility boundary, and **3**: post-fault trajectory does not return to $x_s^{post}$. The single machine infinite bus system is analyzed because it is easy to gain visual insights.

$$\dot{x}_1 = x_2$$
$$M\dot{x}_2 = P_m - \frac{EV}{X}\sin(x_1) - D \times x_2 \quad (13)$$

Here, $x_1$ denotes rotor angle, $x_2$ denotes angular speed deviation, $M$ is inertia, $P_m$ is mechanical power input, $D$ is damping, $E$ is internal emf of the generator, $V$ is voltage of the infinite bus, and $X$ is the total impedance. Fault being analyzed is on the infinite bus i.e. $\frac{EV^{(fault)}}{X} = 0$ and cleared without changing the topology. The constraints assumed are of the form $h(x) = [x_1^{max} - x_1, x_2^{max} - x_2]^T$ arising from out-of-step relay setting for the generator, and frequency threshold from over frequency ride through limit on some large RG in that area. The fixed parameter values are $D = 0.5, \frac{EV^{(pre)}}{X} = \frac{EV^{(post)}}{X} = 1$. Sensitivities are computed at various parameter value combinations, where $p = [P_m, M, \delta_{max}, \omega_{max}]$.

Let us first see the effect of generator mechanical input $P_m$ on CCT at a given operating point. This is important to study as it represents the change in dispatch. The loss of stability at the study point stays the same with fault trajectory intersecting the feasibility boundary under parameter variations. Figure 2 shows the actual CCT vs $P_m$ obtained through TDS. Also shown by dotted lines are CCT estimates at each circled point of the same color using sensitivity formula derived in Section III. C. It can be seen that the dotted lines are tangential to the original curve, which proves the validity of the formula.

Next, we try to understand the implications of changing inertia on meeting frequency constraint $x_2^{max}$. It can be seen from Figure 3 that as the inertia increases, the fault needs to be sustained longer to violate the frequency limits. The trend stays the same up to a certain extent but suddenly changes due to a change in instability phenomenon. Here, for $M \in [0.1: 0.2]$, the sensitivity is calculated using the derivation in Section III. C. while for $M \in [0.25, 0.3]$, it is computed using Section III. E. Again, the sensitivity estimates are tangential to the CCT vs parameter curve, thus validating the formula.

The SRs for the post-fault constrained system plotted under inertia variation is shown in Figure 4. The dark black arrow shows the sustained fault trajectory in each case. Inertia plays the role of reducing the effect of angular excursion on acceleration which can be seen from the changing shape of SRs. For higher inertia values, more angle deviation is needed for the same speed to stabilize. For the given constrained systems, we can see that the stability boundary is comprised of stable manifolds of two semi-saddle pseudo EPs, one on each side of the feasibility boundary, as well as the adjacent unstable portions of the feasibility boundary. These are marked in bold blue and orange, respectively. As the inertia is increased from 0.2 to 0.3, the critical sustained fault trajectory that was earlier resulting in violation of frequency constraints, switches to violating the angle constraint in the post-fault phase. This was because for $M = 0.2$, the exit point $x_{cr}$ of the critical fault trajectory is very close to the intersection of two distinct portions of the stability boundary (horizontal dotted feasibility boundary and orange stable manifold) resulting in a discrete change in the sensitivity function due to small changes in parameter values.

It must also be pointed out that in conventional unconstrained power systems, the parameters under study usually impact the overall system dynamics meaning fault trajectory and all portions of the post fault system stability boundary combined. This makes the relevant portion of the stability boundary structurally stable and consequently the CUEP smoothly varying with parameter changes. As for the constrained systems, the parameters that determine the constraints only impact one or more portions of the feasibility boundary and not the system dynamics itself. This means that with those parameters, the combined stability boundary of the constrained system does not vary and only portions related to the constraints vary. This makes the relevant portion of the feasibility boundary more prone to structural changes as seen in the previous case.

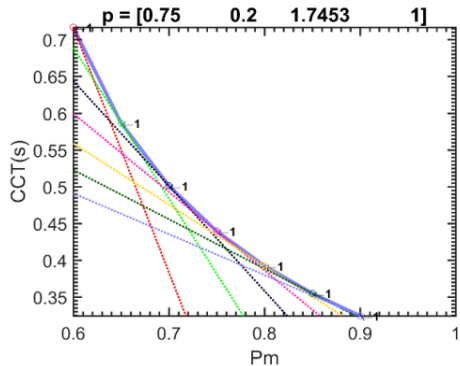

Figure 2 CCT vs $P_m$

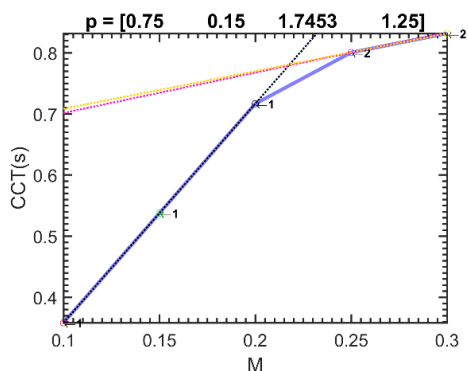

Figure 3 CCT vs M

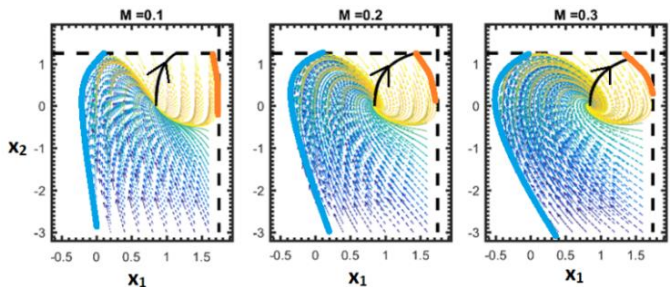

Figure 4 Changing SR with M

## VI. CONCLUSIONS

In this paper, given a critical fault-on and post-fault trajectory, we derived a formula for sensitivity of CCT to parameter variations for systems with inequality constraints. There are multiple instability mechanisms for such systems requiring a sensitivity formula derivation for each. A good application of this could be knowledge of the approximate impact of various system protection settings and operating conditions on changes in likelihood of undesirable tripping without using brute force methods.

It was observed that for constrained systems, the relevant stability boundary may not be structurally stable under parameter variations unlike unconstrained systems. This would require a more sophisticated approach to approximating CCT changes with parameter variations. In this work, we assumed the system had no algebraic constraints. Dobson [8] proposed approaching this problem by converting the DAE system to an ODE system by eliminating the algebraic variable possible due to implicit function theorem. However, this is only possible when the algebraic constraint results in a non-singular Jacobian which is not always the case. This will be explored in a future work.